\documentstyle[psfig,11pt]{article}

\newcommand{\text}[1]{\mbox{#1}}

\begin{document}

\author{Tim Bolton \\
Kansas State University\\
Manhattan, KS 66506-2601}
\title{High Energy Muon Momentum Estimation from Multiple Coulomb Scattering in
Dense Detectors}
\date{May 6, 1997}
\maketitle

\begin{abstract}
{}A method is described for estimating muon track momentum from the
distribution of hits along tracks in dense calorimeters due to multiple
Coulomb scattering (MCS). The method requires only well-aligned conventional
drift chambers or similar tracking devices and can be implemented with or
without a magnetic field present. Neutrino experiments can use this
technique to increase acceptance for large angle charged-current scattering
events. Resolutions in a typical detector geometry vary from $\sim 10\%$ at $%
p=20$ GeV$/c$ to $\sim 50\%$ for $p=500$ GeV$/c$, if the number of hits on
the track is sufficient.
\end{abstract}

\section{ Introduction}

Neutrino detectors are commonly constructed as long dense calorimeters to
maximize interaction rate. This geometry results in loss of acceptance for
charged current $\nu _\mu $ interactions from muons that exit the sides
before they reach the spectrometer that is typically immediately downstream
of the calorimeter. In toroidally magnetized calorimeters, muons can exit
before sufficient $BdL$ is accumulated to measure momentum, or they may
leave a large fraction of their track length in the central hole of the
toroid. Losses are greatest at high values of Bjorken scaling variable $x$
and inelasticity $y$.

Fortunately, these detectors are often instrumented with a large number of
tracking chambers to determined the neutrino interaction vertex and muon
scattering angle. Resolution on this angle can be dominated by multiple
Coulomb scattering (MCS) up to TeV energies. Strong dependence of MCS error
contribution on momentum and the large number of hits on a track in neutrino
detectors permits a different momentum determination scheme. The procedure
dates from the late thirties\cite{EJWilliams}, has been used in many
emulsion experiments\cite{Rossi}, and is still used in balloon-borne cosmic
ray experiments with a variety of tracking technologies\cite{Bertsch}. It
entails a straight line fit to a muon track that varies slope, intercept,
and momentum such that the probability distribution for the observed pattern
of hits is maximized. The MCS-based momentum estimation does not require a
magnetic field and allows for a substantial recovery of the acceptance loss
from exiting muons. Reasonable resolution can be obtained for muons with
momenta up to several hundred GeV$/c$ using a straightforward track finding
and fitting algorithm

The following sections describe the procedure in more detail and the results
of calculations for a detector geometry consisting of $N$ identical tracking
chambers with spatial resolution $\sigma _0$ separated from each other by a
constant thickness $\Delta $ of material with radiation length $X_0$. The
calculations are tested with a Geant\cite{Geant} Monte Carlo simulation of
the NuTeV neutrino experiment\cite{NuTeV} at Fermi National Accelerator
Laboratory. This experiment, chosen for its ``typical'' neutrino detector,
is briefly described in Appendix \ref{NuTeV Detector}. It has parameter
values of $N\leq 42$, $\sigma _0=0.05$ cm, $\Delta =42.4$ cm, and $X_0=3.45$
cm for the purposes of this paper. A forthcoming publication will provide
results of application of the procedure to NuTeV\ data.

\section{Tracks in a Dense Detector}

\subsection{$\chi ^2$ Based Momentum Estimation}

Consider fitting a small-angle muon track to a straight line in a dense
calorimeter instrumented with many equally spaced tracking detectors
(assumed to be drift chambers for the sake of discussion). This may be
accomplished by minimizing a $\chi ^2$ function that compares measured hit
positions to a linear trajectory, 
\begin{equation}
\chi ^2=\left( \vec{y}-\theta _0\vec{z}_1-y_0\vec{z}_0\right) {\bf V}%
^{-1}(p)\left( \vec{y}-\theta _0\vec{z}_1-y_0\vec{z}_0\right) ,
\label{chisq definition}
\end{equation}
with respect to the slope $\theta _0$ and intercept $y_0$. Here, $\vec{y}$
and $\vec{z}_1$ are the $N$ measured $y,z$ points, and $\vec{z}_0$ is an $N$
dimensional vector with all of its elements equal to unity. The covariance
matrix ${\bf V}(p)$ contains constant contributions from chamber resolution
and momentum dependent terms from multiple Coulomb scattering (MCS): 
\begin{equation}
V_{ij}=\sigma _0^2\delta _{ij}+S_{ij}(p),  \label{covariance matrix}
\end{equation}
where the scattering matrix element, usually attributed to Fermi\cite{Rossi}%
, is 
\begin{equation}
S_{ij}(p)=\sum_{k=1}^{\min (i,j)}\frac{\mu _k^2}{p_k^2}\left[ \frac{\Delta
_k^2}3+\frac{\Delta _k}2(z_i-z_k+z_j-z_k)+(z_i-z_k)(z_j-z_k)\right] .
\label{MCS matrix}
\end{equation}
In these expressions for the covariance matrix, $\sigma _0$ is the drift
chamber resolution, $\Delta _k$ is the distance in $z$ between hits $k$ and $%
k-1$, $z_i$ is the distance from the track start to the $i^{th}$ hit, and $%
p_k$ is the momentum (in GeV$/c$) in the gap between hit $k$ and $k-1$; $\mu
_k\simeq 0.015\sqrt{\Delta _k/X_k}$, with $X_k$ the radiation length, is a
constant depending on the composition and thickness of the tracking medium.
Parametrizations for $\mu _k$ are discussed further in Appendix \ref{MCS
parameter}. The rms displacement in the length $\Delta _k$ of 
\begin{equation}
\delta _k=\sqrt{\frac 13}\frac{\mu _k\Delta _k}{p_k},
\end{equation}
is, for iron, given by 
\begin{equation}
\delta _k\simeq 320\mbox{ }\mu \mbox{m }\frac{10\mbox{ GeV}/c}p\left( \frac{%
\Delta _k}{10\mbox{ cm}}\right) ^{3/2}.
\end{equation}
For 10 cm tracking chamber separation, this displacement is the same as a
typical spatial resolution measurement of a drift chamber.

For constant chamber separation, one can set $\Delta _k=\Delta $, $\mu
_k=\mu $ and incorporate energy loss effects in an approximate way to yield
simplification: 
\begin{eqnarray}
S_{ij}(p) &\simeq &\frac{\mu ^2\Delta ^2\min \left( i,j\right) }{6p^2}%
\left\{ \left[ 2\min \left( i,j\right) ^2-3\left( i+j\right) \min \left(
i,j\right) +6ij\right] \right. \\
&&+\frac \Delta p\left\langle \frac{dp}{dz}\right\rangle \left[ \left( \min
\left( i,j\right) +1\right) 3\min \left( i,j\right) ^2-4\left( i+j\right)
\min \left( i,j\right) \right.  \nonumber \\
&&\left. \left. +i+j-\min \left( i,j\right) \right] \right\} ,
\end{eqnarray}
with $p$ the momentum at the start of the track. Many of the formulas
presented here will assume the mean energy loss, $\left\langle \frac{dp}{dz}%
\right\rangle $, is zero for simplicity, although, as will be seen,
incorporation of finite $\left\langle \frac{dp}{dz}\right\rangle $ can
significantly improve momentum estimates from MCS.\footnote{%
For very low $p$ tracks in long targets, muons can range out, in which case
momentum determination from $\left\langle \frac{dp}{dz}\right\rangle $ is
possible, {\it in addition to} the MCS estimate.}

If the calorimeter is instrumented with a sufficiently large number of drift
chambers, it is possible to exploit MCS to estimate muon track momentum from
the scatter of the hits along a muon track. This can be seen from the
following intuitive argument: Best estimates for $\theta $ and $y_0$ follow
from minimizing the $\chi ^2:$%
\begin{eqnarray}
\theta _0 &=&\frac{<yz_1>-<yz_0><z_1z_0>}{<z_1z_1>-<z_1z_0>^2},
\label{slope parameter} \\
y_0 &=&\frac{<z_1z_1><yz_0>-<z_1z_0><yz_1>}{<z_1z_1>-<z_1z_0>^2}.
\label{intercept parameter}
\end{eqnarray}
Bracketed quantities $<ab>$ are defined as 
\begin{equation}
<ab>=\frac{\vec{a}{\bf V}^{-1}\vec{b}}{\vec{z}_0{\bf V}^{-1}\vec{z}_0};
\label{reduced matrix}
\end{equation}
they are unchanged by a re-scaling of the error matrix. In the MCS limit, it
follows that $\theta _0$ and $y_0$ are independent of the momentum, implying
that $\chi ^2\propto p^2$. Now, suppose one adjusts $p$ until the $\chi ^2$
probability density function attains its maximum. This occurs at $\chi
^2\simeq N$, and if the fit is performed at some nominal momentum $p_0$
achieving $\chi ^2=\chi _0^2$, then an estimate for the true momentum of the
track is 
\begin{equation}
p=p_0\sqrt{\frac N{\chi _0^2}.}  \label{p estimate}
\end{equation}
The MCS\ technique thus provides a momentum estimation method that does not
require a magnetic field.

\subsection{Likelihood Function Method}

A more rigorous derivation begins with the observation that the joint
probability function for $N$ correlated drift chamber hits can be written,
assuming Gaussian errors, as 
\begin{equation}
P(\vec{y};\theta _0,y_0,p)=\left( \frac 1{2\pi }\right) ^{N/2}\frac 1{\sqrt{%
\det {\bf V}(p)}}\exp \left[ -\frac 12\chi ^2(\theta _0,y_0,p)\right] ,
\end{equation}
with $\chi ^2$ defined in equation \ref{chisq definition}. This can be
converted to a $(-)$log likelihood function, 
\begin{equation}
{\cal L}=\frac 12\log (\det {\bf V}(p))+\frac 12\chi ^2(\theta _0,y_0,p),
\end{equation}
where terms independent of $y_0$, $\theta _0$, and $p$ have been dropped.
Estimates for $y_0$, $\theta _0$, and $p$ follow from minimizing ${\cal L}$.

\subsubsection{MCS dominated limit}

In the multiple scattering limit, one can write ${\bf V}(p)\simeq \frac{p_0^2%
}{p^2}{\bf V}(p_0)$, where $p_0$ is some nominal estimate of the momentum.
From this, it follows that $\det {\bf V}(p)=\left( \frac{p_0^2}{p^2}\right)
^N\det {\bf V}(p_0)$ and $\chi ^2(\theta _0,y_0,p)=\frac{p^2}{p_0^2}\chi
^2(\theta _0,y_0,p_0)$. In this limit, the log likelihood becomes, after
dropping terms that are independent of $\theta _0$, $y_0$, and $p$%
\begin{equation}
{\cal L}\rightarrow -N\log p+\frac{p^2}{2p_0^2}\chi ^2(\theta _0,y_0,p_0).
\end{equation}
Minimizing with respect to the three fit parameters yields $\frac{\partial
\chi ^2}{\partial \theta _0}=\frac{\partial \chi ^2}{\partial y_0}=0$, as
before, and equation \ref{p estimate}. One can also obtain an estimate of
the uncertainty in $p$ from 
\begin{equation}
\frac 1{\sigma _p^2}=\frac{\partial ^2{\cal L}}{\partial p^2}_{{\cal L=L}%
_{\max }}=\frac{2\chi ^2(\theta _0,y_0,p_0)}{p_0^2}.
\end{equation}
If the fit is iterated until $p=p_0$ and the fit is reasonable so that $\chi
^2\simeq N$, then 
\begin{equation}
\frac{\sigma _p}p\rightarrow \frac 1{\sqrt{2N}}\mbox{ (MCS limit)}.
\label{MCS limit error}
\end{equation}

\subsubsection{Effects of Finite Spatial Resolution}

In the more typical case where chamber resolution is not negligible, one
must solve the coupled equations 
\begin{eqnarray}
\frac \partial {\partial y_0}\chi ^2(\theta _0,y_0,p) &=&0
\label{y-equation} \\
\frac \partial {\partial \theta _0}\chi ^2(\theta _0,y_0,p) &=&0,
\label{theta-equation} \\
\frac \partial {\partial p}\left[ \log (\det {\bf V}(p))+\chi ^2(\theta
_0,y_0,p)\right] &=&0,  \label{p-equation}
\end{eqnarray}
which can be accomplished via straightforward iterative methods by computer.

\begin{figure}[pthb]
\psfig{file=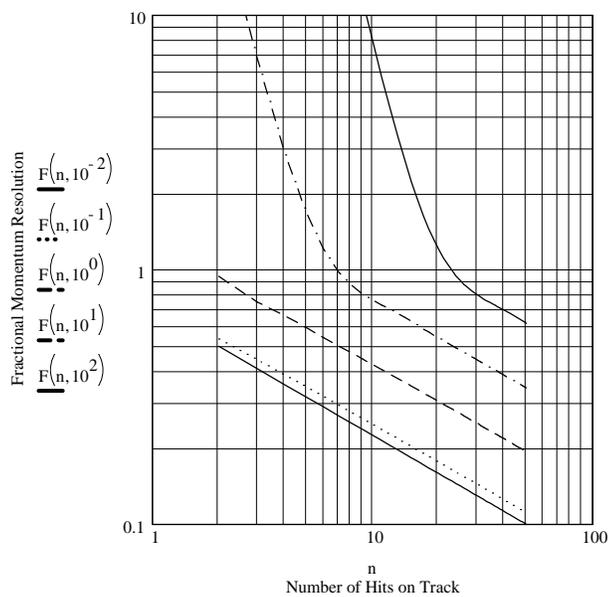,clip=,height=3.5in}
\caption{ Universal resolution function $F(n,x)$ for MCS determination of
momentum. The curves represent different values of $x=\frac{p\sigma_0}{%
\mu\Delta}$ which are $x=0.01$ (solid-lower), $x=0.1$ (dots), $x=1$
(dashed), $x=10$ (dot-dashed), and $x=100$ (solid-top). }
\label{universal}
\end{figure}

\begin{figure}[pthb]
\psfig{file=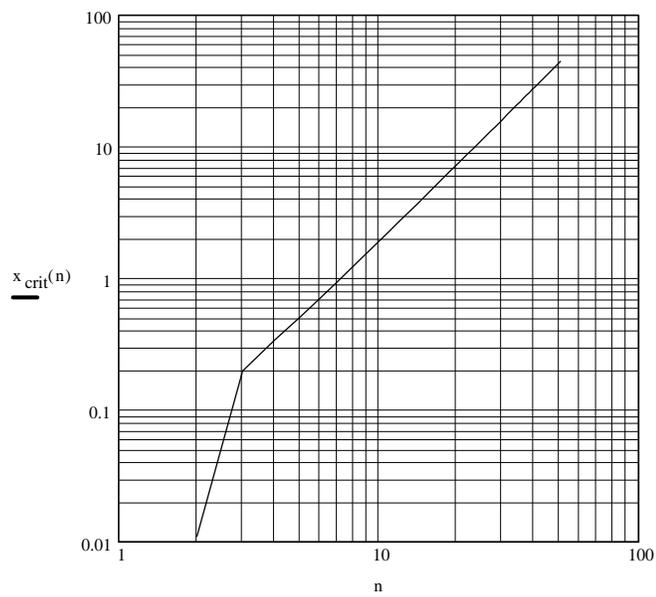,clip=,height=3.5in}
\caption{ Critical value of $x=\frac{p\sigma_0}{\mu\Delta}$ as a function of
number of drift chamber hits. For this value of $x$, the fractional momentum
resolution will be 50\%. }
\label{xcrit}
\end{figure}

\begin{figure}[pthb]
\psfig{file=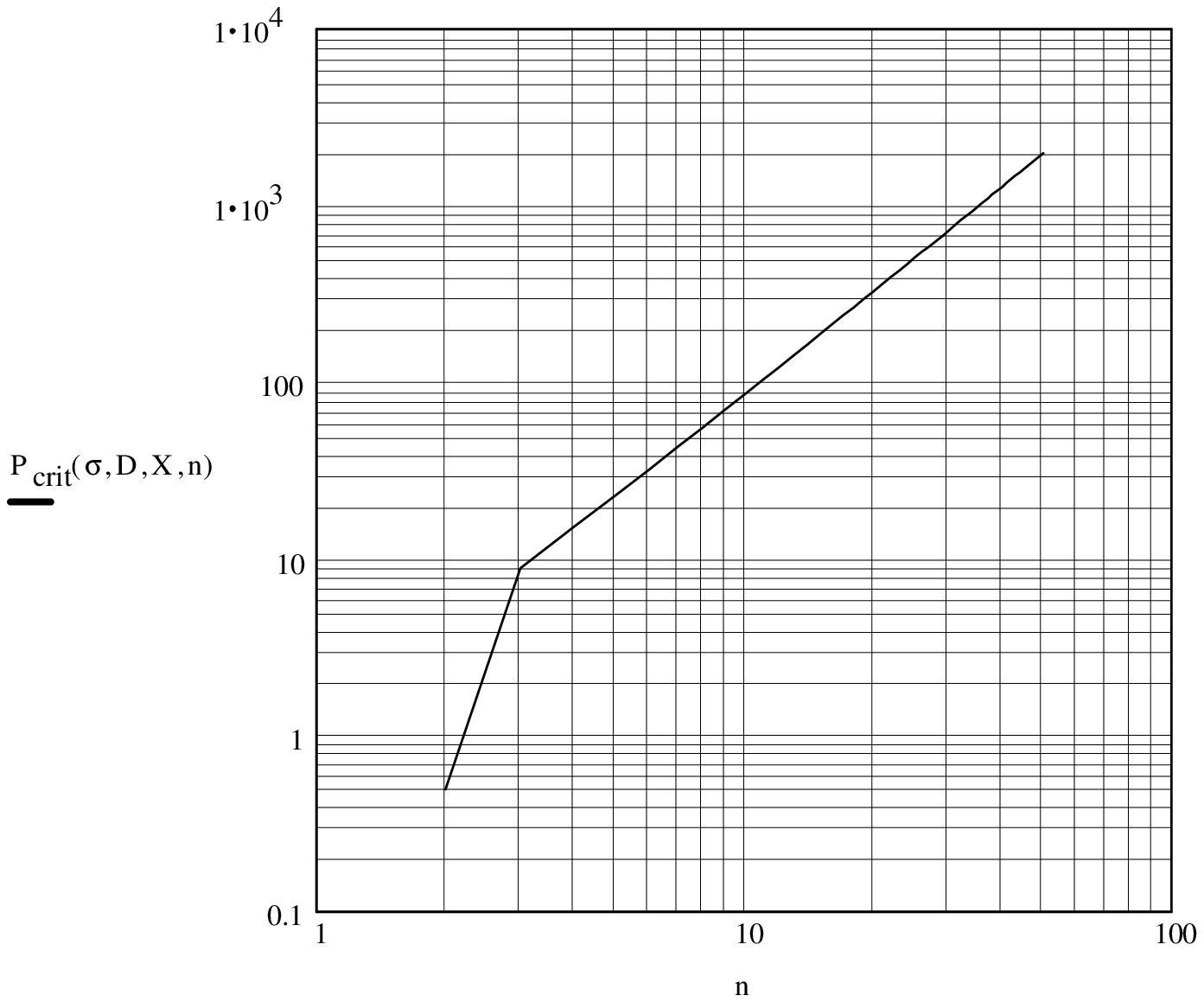,clip=,height=3.5in}
\caption{ Value of critical momentum $P_{crit}$, above which the fractional
momentum resolution will exceed 50\% for a given number of drift chambers $n$%
. This plot assumes the NuTeV detector geometry, with $\sigma_0=0.05$ cm, $%
\Delta=42.4$ cm and 12.2 radiation lengths of material between each
tracking chamber. }
\label{pcrit}
\end{figure}

Insight into the intrinsic resolution of the MCS\ momentum error estimate
can be gained by examining an approximate expression for the momentum
resolution, derived in Appendix \ref{Resolution}: 
\begin{equation}
\frac{\sigma _p}p=\left[ 2\sum_{n=1}^N\frac{\xi _n^4}{\left( \xi _n^2+\frac{%
p^2\sigma _0^2}{\mu ^2\Delta ^2}\right) ^2}\right] ^{-1/2},
\label{MCS P-error}
\end{equation}
where $\xi _n^2$ are the eigenvalues of the dimensionless scattering matrix $%
{\bf \tilde{S}}=\frac{p^2}{\mu ^2\Delta ^2}{\bf S}(p)$.\footnote{%
This can be expressed in the computationally simpler form $p^2/\sigma
_p^2=2Tr\left[ {\bf V}(p)^{-2}{\bf S}^2(p)\right] $.} For the geometry
considered here, and ignoring energy loss, the relative momentum error is
seen to be a universal function of the number of chambers $n$ and the ratio $%
p^2/p_{MCS}^2,$ 
\begin{equation}
\frac{\sigma _p}p=F(n,p^2/p_{MCS}^2),
\end{equation}
where 
\begin{equation}
p_{MCS}=\frac{\mu \Delta }{\sigma _0},
\end{equation}
defines a characteristic momentum scale (approximately $73$ GeV\ for the
NuTeV\ detector). Figure \ref{universal} shows plots of $F(n,x)$ vs $n$ for
different values of $x=p^2/p_{MCS}^2$. About 7 chambers are required to
measure $p_{MCS}$ to $50\%$ fractional momentum error and 25 chambers to
measure $10\times p_{MCS}$ to $50\%$. For a given number of chambers $n$
used on a track fit, one can define a critical value $x_{crit}(n)$, such
that $\sigma _p/p\leq 50\%$ for $x\leq x_{crit}(n)$. Figure \ref{xcrit}
shows a plot of $x_{crit}(n)$ vs $n$. For a given detector geometry, $%
x_{crit}(n)$ can be converted to $p_{crit}(n)$, the largest momentum that
can be estimated from MCS scattering alone to $50\%$ resolution. Figure \ref
{pcrit} shows a plot of $p_{crit}(n)$ vs $n$ for the NuTeV\ detector.
Momentum values of to 300 GeV can be estimated using 21 chambers in the
detector, and up to 1 TeV using all 42 chambers.

The MCS limit is $p^2/p_{MCS}^2\ll \xi _n^2$ for all $n$, in which case Eq. 
\ref{MCS limit error} is recovered. If intrinsic chamber resolution
dominates, MCS-based momentum estimates will then provide resolutions that
behave as 
\begin{equation}
\frac{\sigma _p}p\rightarrow \sqrt{\frac 1{2N}}\left( \frac
p{p_{MCS}}\right) ^2\mbox{ (resolution limit),}.
\end{equation}

\subsection{Tracking in a Magnetic Field}

The much more straightforward way to determine momentum is via track
displacement in a magnetic field. It is possible to combine the curvature
measurement with the MCS momentum technique to improve the overall momentum
estimation.

For simplicity, the analysis will be restricted to a geometry of evenly
spaced tracking chambers immersed in a uniform magnetic field oriented at
right angles to the track propagation. It is further assumed that the magnet 
$p_T$ kick is much less that the momentum of the track being analyzed. In
this case there is no dependence of the covariance matrix on fit parameters
other than momentum, and the variance matrix for the fitted momentum takes
the form 
\begin{equation}
{\bf E}_{pp}^{-1}={\bf \Psi }^{-1}+\frac 1{\sigma _p^2}\left( 
\begin{array}{lll}
0 & 0 & 0 \\ 
0 & 0 & 0 \\ 
0 & 0 & 1
\end{array}
\right) ,
\end{equation}
where $\sigma _p^2$ is given Eq. \ref{MCS P-error} and ${\bf \Psi }$ is the
conventional spectrometer error matrix, with 
\begin{eqnarray}
\Psi _{11}^{-1} &=&\vec{z}_0{\bf V}^{-1}(p)\vec{z}_0, \\
\Psi _{12}^{-1} &=&\vec{z}_1{\bf V}^{-1}(p)\vec{z}_0, \\
\Psi _{22}^{-1} &=&\vec{z}_1{\bf V}^{-1}(p)\vec{z}_1, \\
\Psi _{13}^{-1} &=&-\frac k{2p^2}\vec{z}_0{\bf V}^{-1}(p)\vec{z}_2, \\
\Psi _{23}^{-1} &=&-\frac k{2p^2}\vec{z}_1{\bf V}^{-1}(p)\vec{z}_2, \\
\Psi _{33}^{-1} &=&\frac{k^2}{4p^4}\vec{z}_2{\bf V}^{-1}(p)\vec{z}_2,
\end{eqnarray}
Here, 
\begin{eqnarray}
\vec{z}_0 &=&\left( 1,1,1,...,1\right) , \\
\vec{z}_1 &=&\left( z_1,z_2,z_3,...,z_N\right) , \\
\vec{z}_2 &=&\left( z_1^2,z_2^2,z_3^2,...,z_N^2\right) ,
\end{eqnarray}
and $k=0.003B$, with $B$ the magnetic field in Tesla assuming all spatial
coordinates are in cm.

In some detectors, such as NuTeV, the spectrometer follows the calorimeter.
Spectrometer momentum determination and MCS-based calorimeter determination
are then independent and can be averaged.

\section{Results from Calculations}

\subsection{Tracking in Unmagnetized Calorimeter}

\begin{figure}[pthb]
\psfig{file=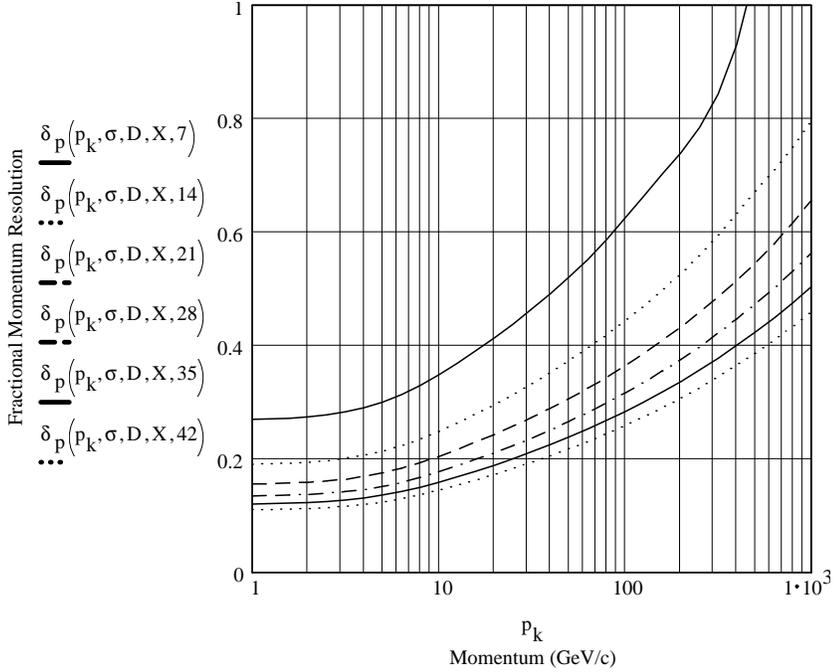,clip=,height=3.5in}
\caption{Momentum dependence of fractional momentum resolution for different
numbers of chambers used in the fit. The curves correspond to 7 chambers
(upper-solid), 14 chambers (upper-dotted), 21 chambers (dashed), 28 chambers
(dot-dashed), 35 chambers (lower-solid), and the maximum possible 42
chambers (lower-dotted). This plot assumes the NuTeV detector geometry,
with $\sigma_0=0.05$ cm, $\Delta=42.4$ cm, and 12.2 radiation lengths of
material between each tracking chamber. }
\label{momentum calc}
\end{figure}

\begin{figure}[pthb]
\psfig{file=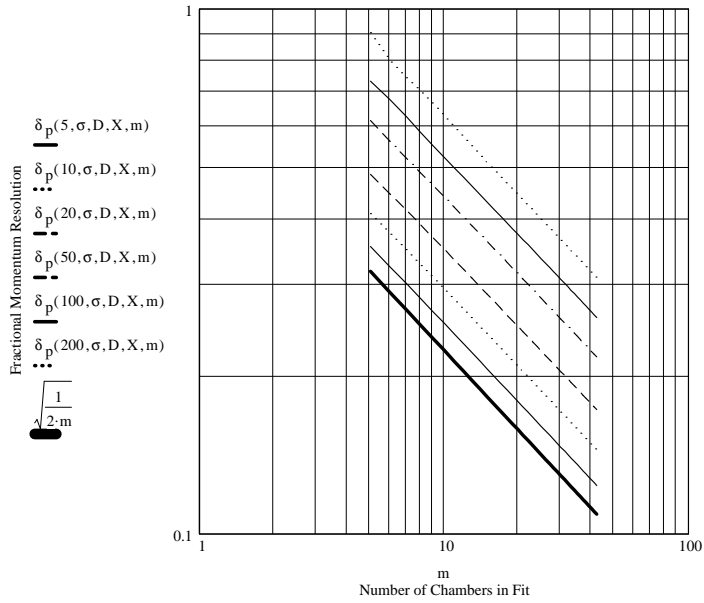,clip=,height=3.5in}
\caption{Dependence on number of drift chamber hits of fractional momentum
resolution for different muon momenta. The dark solid curve represents the
MCS limit. The other curves correspond to $p=5$ GeV$/c$ (lower-lighter
solid), $p=10$ GeV$/c$ (lower-dotted), $p=20$ GeV$/c$ (dashed), $p=50$ GeV$%
/c $ (dot-dashed), $p=100$ GeV$/c$ (upper-solid), and $p=200$ GeV$/c$
(upper-dotted). This plot assumes the NuTeV detector geometry, with $%
\sigma_0=0.05$ cm, $\Delta=42.4$ cm, and 12.2 radiation lengths of material
between each tracking chamber. }
\label{hits calc}
\end{figure}

\begin{figure}[pthb]
\psfig{file=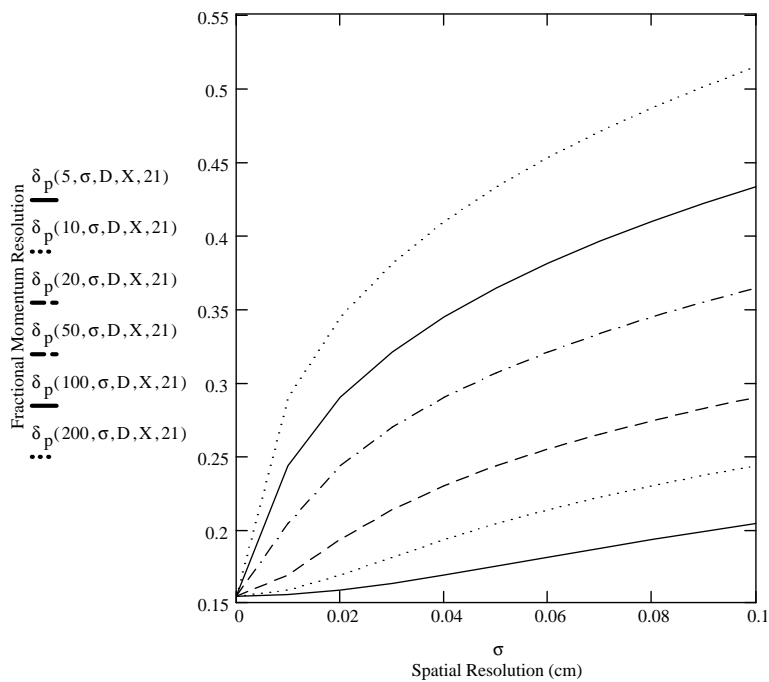,clip=,height=3.5in}
\caption{Spatial resolution dependence of fractional momentum resolution for
different muon momenta assumes 21 drift chamber hits in the NuTeV geometry.
The curves correspond to $p=5$ GeV$/c$ (lower-solid), $p=10$ GeV$/c$
(lower-dotted), $p=20$ GeV$/c$ (dashed), $p=50$ GeV$/c$ (dot-dashed), $p=100$
GeV$/c$ (upper-solid), and $p=200$ GeV$/c$ (upper-dotted). This plot
otherwise assumes the NuTeV detector geometry, with $\Delta=42.4$ cm, and
12.2 radiation lengths of material between each tracking chamber. }
\label{sigma calc}
\end{figure}

Figures \ref{momentum calc}, \ref{hits calc}, and \ref{sigma calc} show
results for the estimated fractional momentum error $\delta _P=(\sigma _p/p)$
calculated for the NuTeV detector geometry from Eq. \ref{MCS P-error} as a
function of various parameters appearing in Eq. \ref{MCS P-error}.

Figure \ref{momentum calc} shows the dependence of fractional resolution on
momentum for various numbers of drift chambers. Momentum dependence is
present for all momenta and all numbers of chambers, indicating that the MCS
resolution limit of $\sigma _p/p=1/\sqrt{2N}$ is not reached until lower
momentum.

Figure \ref{hits calc} shows the dependence of $\sigma _p/p$ on the number
of drift chamber hits. While the $1/\sqrt{2N}$ limit is not reached, the
resolution does scale as $A/\sqrt{2N}$, with $A$ increasing with momentum.

Figure \ref{sigma calc} shows the dependence on chamber resolution. Effects
are sizable, indicating that a careful assessment of the intrinsic chamber
resolution is necessary.

\subsection{Tracking in Magnetized Calorimeter}

\begin{figure}[pthb]
\psfig{file=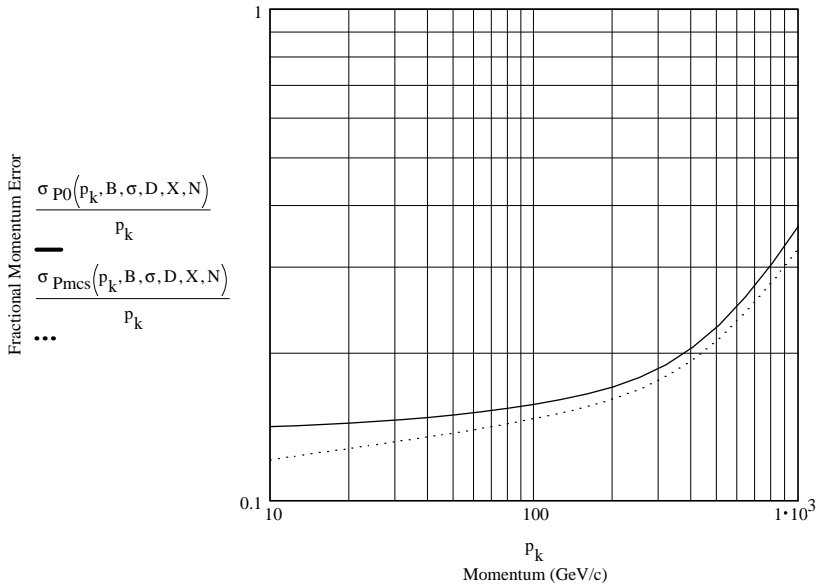,clip=,height=3.5in}
\caption{ Dependence of fractional momentum resolution on muon momentum
assuming twenty 0.05 cm resolution drift chamber hits spaced by 10 radiation
lengths of iron. The top (solid) curve assumes a conventional spectrometer
fit, while the lower (dotted) curve incorporates information from MCS into
the fit.}
\label{B-P-dep}
\end{figure}

\begin{figure}[pthb]
\psfig{file=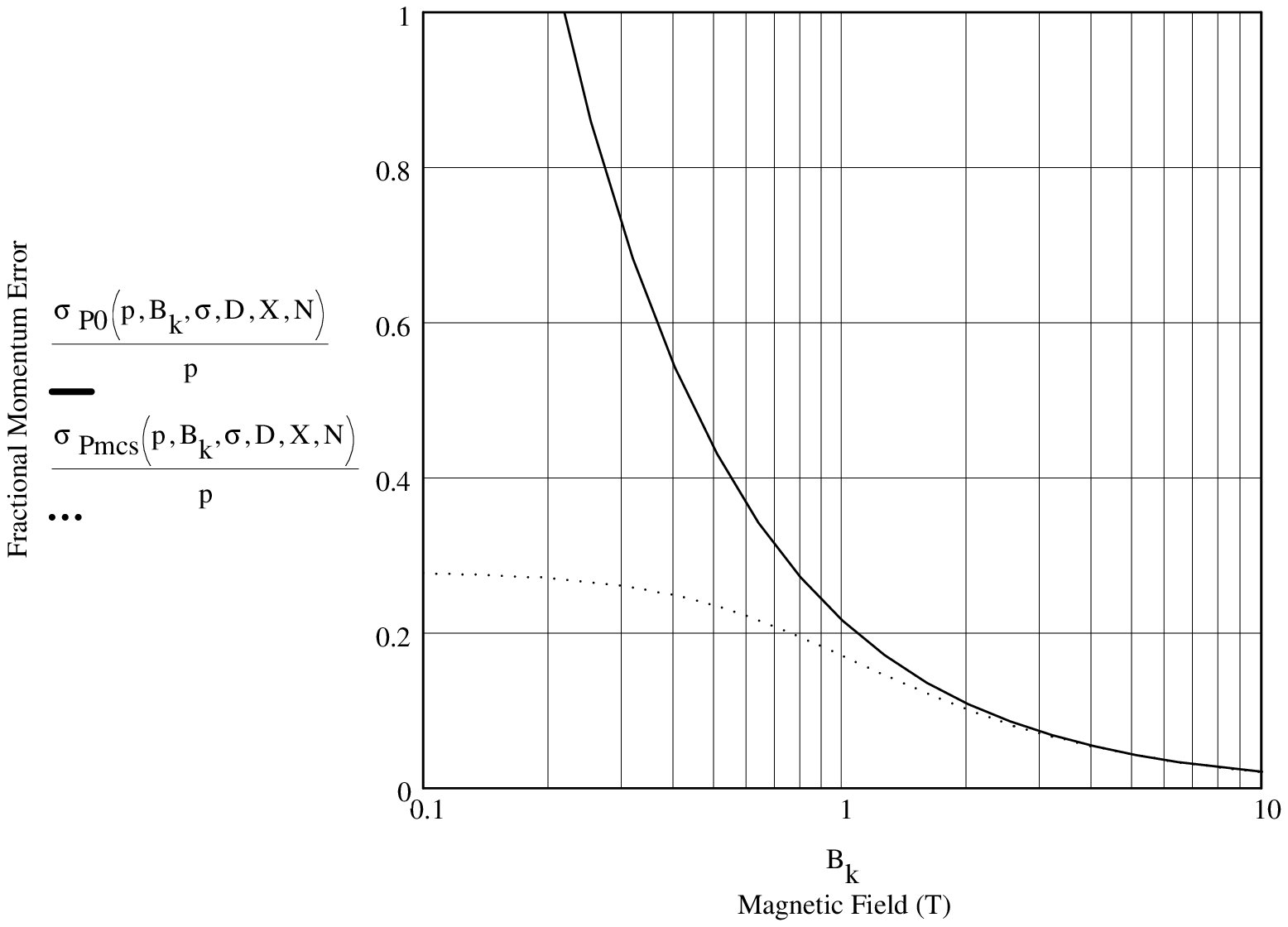,clip=,height=3.5in}
\caption{ Dependence of fractional momentum resolution on magnetic field
assuming twenty 0.05 cmm resolution drift chamber hits spaced by 10
radiation lengths of iron. The top (solid) curve assumes a conventional
spectrometer fit, while the lower (dotted) curve incorporates information
from MCS into the fit.}
\label{B-B-dep}
\end{figure}

\begin{figure}[tbph]
\psfig{file=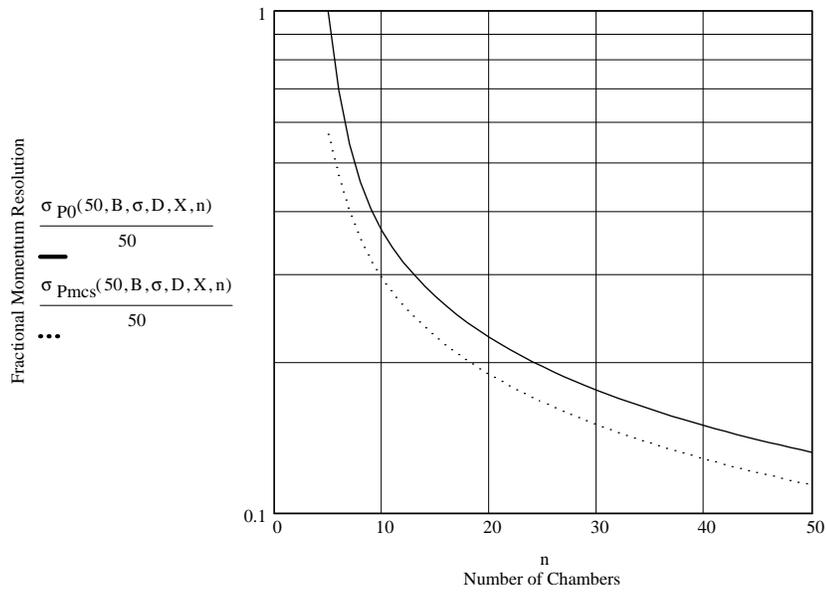,clip=,height=3.5in}
\caption{ Dependence of fractional momentum resolution on number of drift
chamber hits for 50 GeV muon momentum and 0.05 cm resolution drift chambers
spaced by 10 radiation lengths of iron. The top (solid) curve assumes a
conventional spectrometer fit, while the lower (dotted) curve incorporates
information from MCS into the fit.}
\label{B-N-dep}
\end{figure}

Figure \ref{B-P-dep}, \ref{B-B-dep}, and \ref{B-N-dep} show the dependence
of fractional momentum resolution in a magnetized calorimeter as a function
of muon momentum, magnetic field, and number of chambers respectively. Also
shown is the resolution estimate for a conventional momentum fit that
incorporates MCS effects into the error matrix, but uses only the track
curvature, not the pattern of scatter in the hits, to estimate momentum. The
three plots assume a geometry with 0.05 cm resolution drift chambers
separated by 10 radiation lengths of material. Figures \ref{B-P-dep} and \ref
{B-N-dep} assume a magnetic field of 1 T, Figs. \ref{B-B-dep} and \ref
{B-N-dep} assume a muon momentum of 50 GeV, and Figs. \ref{B-B-dep} and \ref
{B-N-dep} assume 20 chambers used in the fit.

\begin{figure}[tbph]
\psfig{file=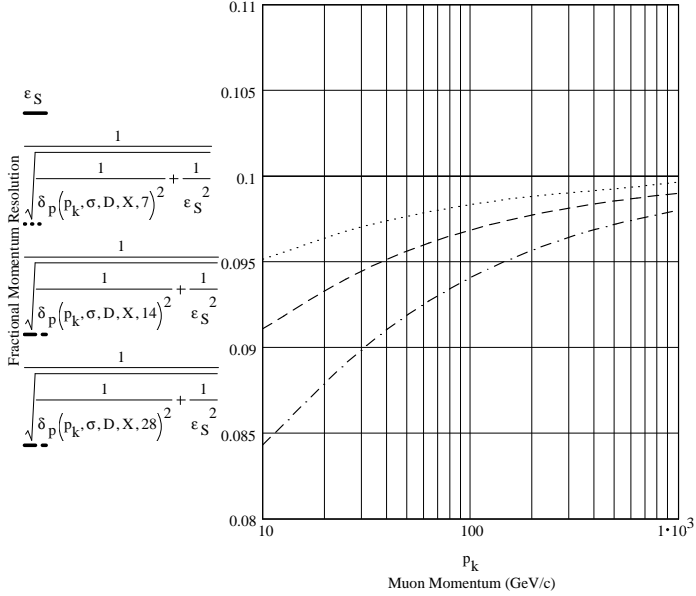,clip=,height=3.5in}
\caption{ Dependence of fractional momentum resolution on muon momentum for
different number of chambers used in the MCS determination in a momentum
estimate that combines the MCS estimate with a 10\% spectrometer
measurement, $\varepsilon_S$, in the NuTeV geometry ($\sigma_0=0.05$ cm, $%
\Delta=42.4$ cm, and 12.2 radiation lengths of material between each
tracking chamber). The curves correspond to 28 chambers (dot-dashed), 14
chambers (dashed), and 7 chambers (dotted). The solid horizontal line at 0.1
represents the spectrometer-only resolution. }
\label{combined resolution}
\end{figure}

Figure \ref{combined resolution} shows the combined momentum resolution that
can be achieved from the spectrometer and a varying number of calorimeter
chambers used in the NuTeV experiment. The spectrometer alone provides a
resolution of $\varepsilon_S=10\%$.

\section{Results of Monte Carlo Simulation}

The formulas developed in the previous section have been tested using a
Geant simulation of the NuTeV\ detector (see Appendix \ref{NuTeV Detector})
using track finding and fitting algorithms described in Appendix \ref{Fit
Procedure}. This section presents results only for tracking in the
unmagnetized NuTeV calorimeter..

\begin{figure}[tbph]
\psfig{file=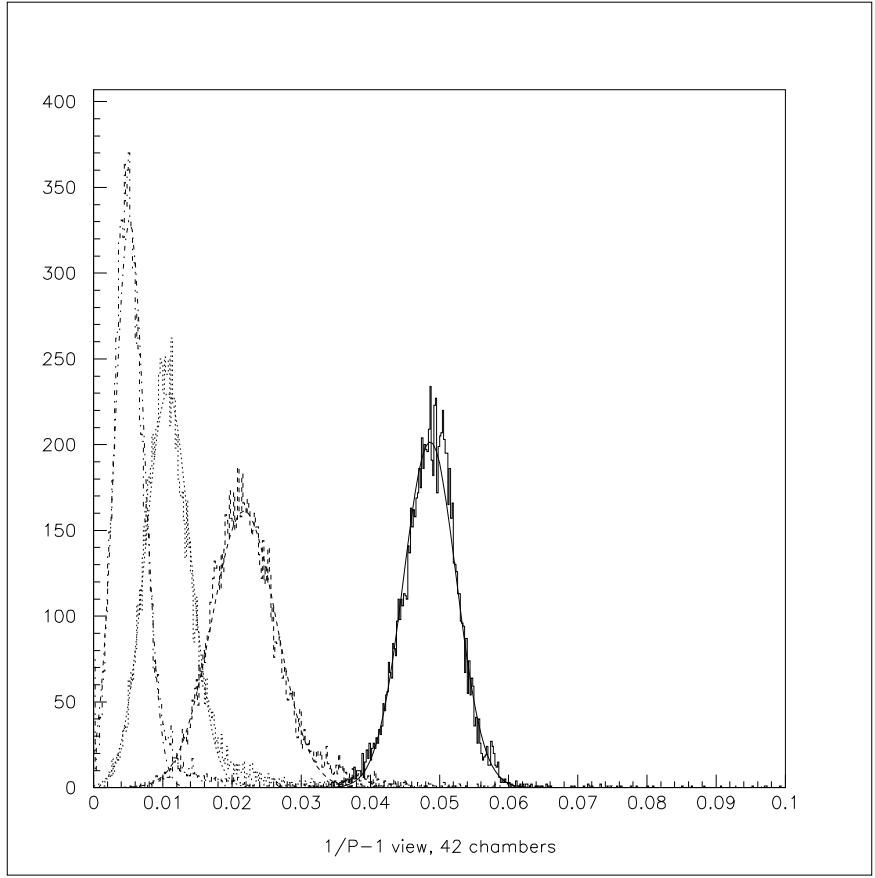,clip=,height=5.0in}
\caption{ Distributions of $1/p$ (in (GeV$/c$)$^{-1}$) estimated from MCS
technique for 20 GeV$/c$, 50 GeV$/c$, 100 GeV$/c$, and 200 GeV$/c$ muons
passing through 42 drift chambers in a Geant simulation of the NuTeV
neutrino detector. Only one of two drift chamber views is used in the
fitting. The histograms are for, from right to left, 20 GeV$/c$ (solid), 50
GeV$/c$ (dashed), 100 GeV$/c$ (dotted), and 200 GeV$/c$ (dot-dashed) muons,
respectively. The curves superimposed on the histograms represent simple
Gaussian fits. }
\label{P-plot1}
\end{figure}

\begin{figure}[tbph]
\psfig{file=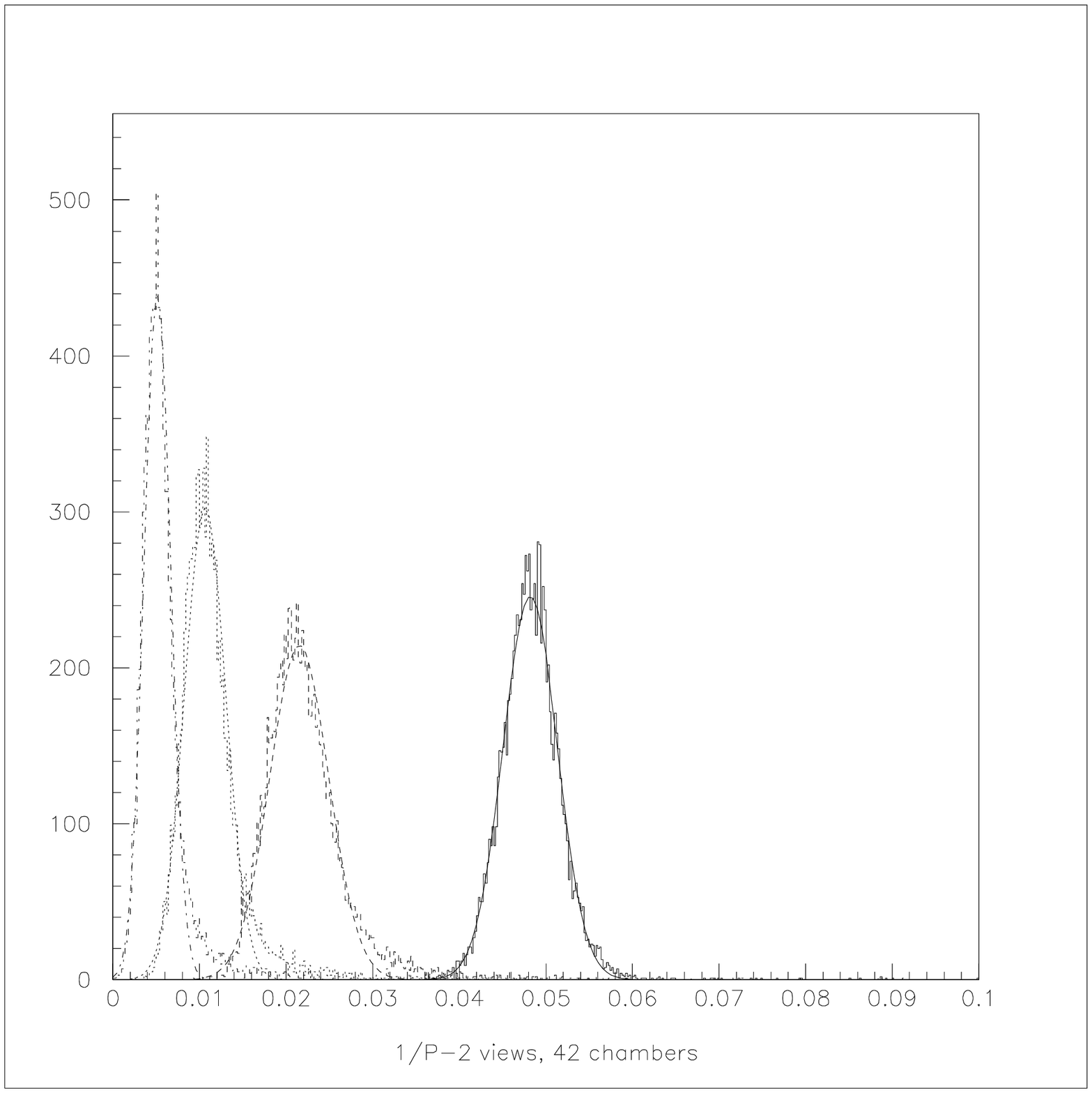,clip=,height=5.0in}
\caption{ Distributions of $1/p$ (in (GeV$/c$)$^{-1}$) estimated from MCS
technique for 20 GeV$/c$, 50 GeV$/c$, 100 GeV$/c$, and 200 GeV$/c$ muons
passing through 42 drift chambers in a Geant simulation of the NuTeV
neutrino detector. Both drift chamber views are used in the fitting. The
histograms are for, from right to left, 20 GeV$/c$ (solid), 50 GeV$/c$
(dashed), 100 GeV$/c$ (dotted), and 200 GeV$/c$ (dot-dashed) muons,
respectively. The curves superimposed on the histograms represent simple
Gaussian fits. }
\label{P-plot2}
\end{figure}

Figures \ref{P-plot1} and \ref{P-plot2} show distributions of fitted values
of $1/p$ as a function of track momentum using all 42 chambers in the NuTeV\
detector. Results are presented for momentum determination using only a
single view in the drift chamber, and for fits that combine both views.
Table \ref{P-table} summarizes momentum dependence of reconstructed
momentum, fractional resolution, and tracking efficiency.

\begin{table}[tbp] \centering%
\begin{tabular}{|l|l|l|l|l|}
\hline
${\bf p}_{in}${\bf \ (GeV}$/c${\bf )} & 20 & 50 & 100 & 200 \\ \hline
${\bf p}_{rec}^{\mbox{1 view}}${\bf (GeV}$/c${\bf )} & 20.6 & 46.0 & 93.4 & 
194.0 \\ \hline
${\bf p}_{rec}^{\mbox{2 views}}${\bf (GeV}$/c${\bf )} & 20.8 & 46.5 & 94.0 & 
195.0 \\ \hline
${\bf \sigma /p}^{\mbox{1 view}}$ & 0.078 & 0.21 & 0.30 & 0.42 \\ \hline
${\bf \sigma /p}^{\mbox{1 view}}${\bf (pred)} & 0.17 & 0.22 & 0.26 & 0.31 \\ 
\hline
${\bf \sigma /p}^{\mbox{2 views}}$ & 0.065 & 0.15 & 0.21 & 0.30 \\ \hline
${\bf \sigma /p}^{\mbox{2 views}}${\bf (pred)} & 0.12 & 0.15 & 0.18 & 0.22
\\ \hline
${\bf \epsilon }_p^{\mbox{1 view}}{\bf (\%)}$ & 99 & 99 & 99 & 96 \\ \hline
${\bf \epsilon }_p^{\mbox{2 views}}{\bf (\%)}$ & 99 & 99 & 99 & 93 \\ \hline
\end{tabular}
\caption{Momentum dependence of MCS momentum estimate
from a Geant simulation of the NuTeV calorimeter. Results for
reconstructed momentum ($p_{rec}$),  resolution ($\sigma_p/p$),
and reconstruction efficiency ($\epsilon_p$)  are given assuming that either one
or two views of drift chamber hits are used. The Monte Carlo sample
contained $10^4$ events, so statistical errors are of order 1\%.
\label{P-table}}%
\end{table}%

\begin{figure}[tbph]
\psfig{file=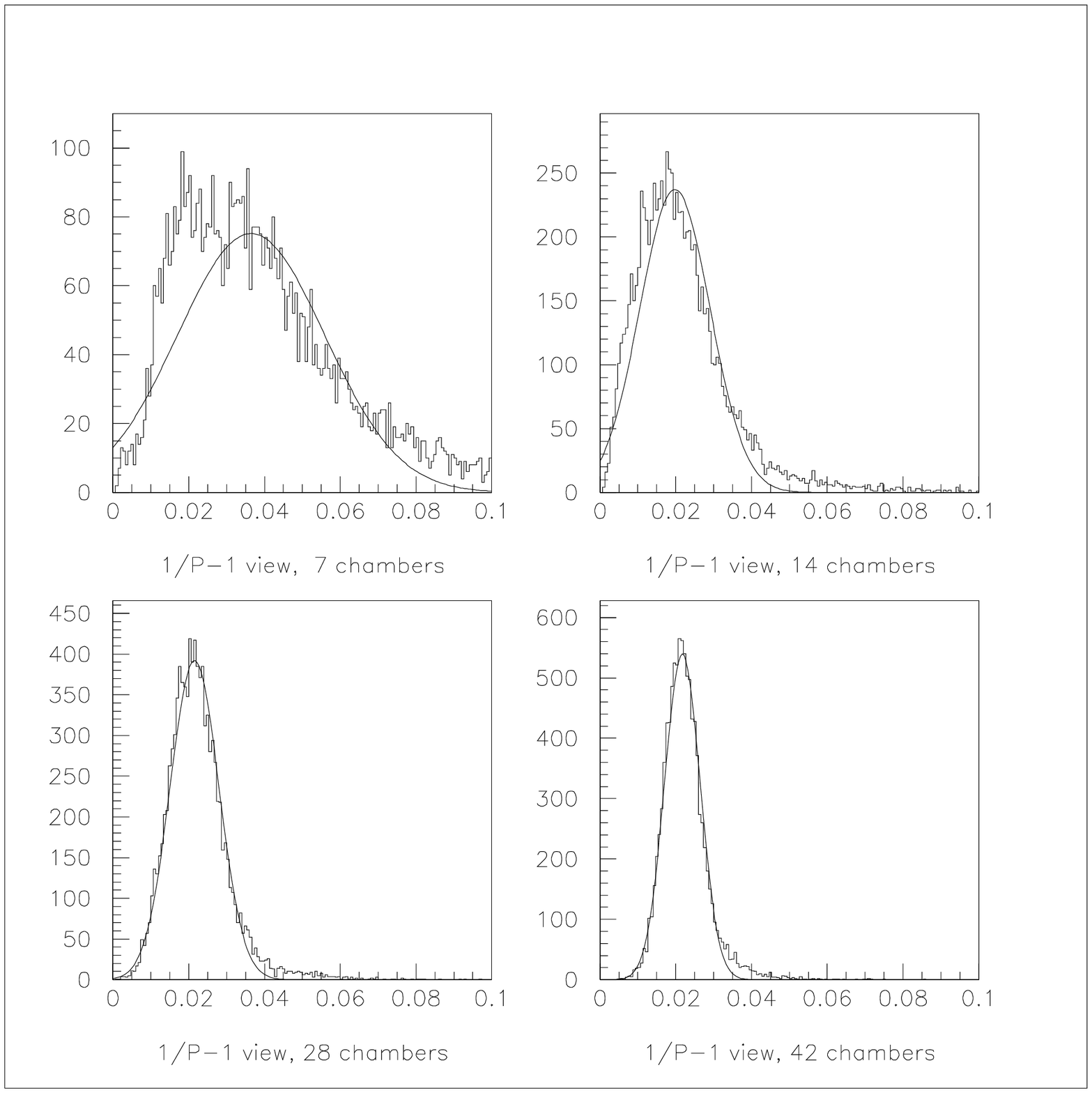,clip=,height=5.0in}
\caption{ Distributions of $1/p$ (in (GeV$/c$)$^{-1}$) estimated from MCS
technique for 50 GeV$/c$ muons passing through different numbers of drift
chambers in a Geant simulation of the NuTeV neutrino detector. Only one of
two drift chamber views is used in the fitting. The plots are for, clockwise
from top-left, 7 chambers, 14 chambers, 42 chambers, and 28 chambers used in
the fit. The superimposed curves represent simple Gaussian fits. }
\label{N-plot1}
\end{figure}

\begin{figure}[tbph]
\psfig{file=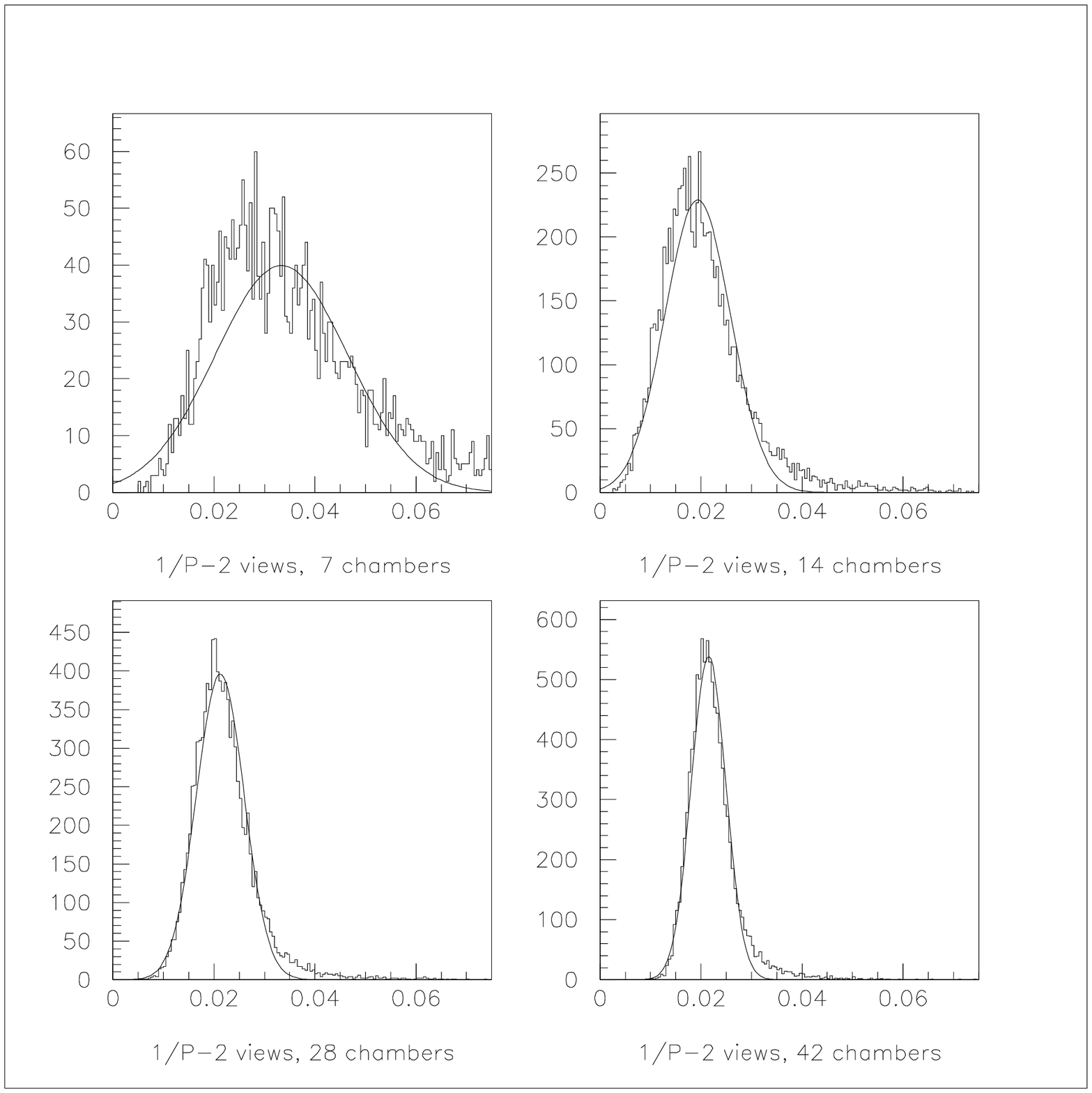,clip=,height=5.0in}
\caption{ Distributions of $1/p$ (in (GeV$/c$)$^{-1}$) estimated from MCS
technique for 50 GeV$/c$ muons passing through different numbers of drift
chambers in a Geant simulation of the NuTeV neutrino detector. Both drift
chamber views are used in the fitting. The plots are for, clockwise from
top-left, 7 chambers, 14 chambers, 42 chambers, and 28 chambers used in the
fit. The curves superimposed on the histograms represent simple Gaussian
fits. }
\label{N-plot2}
\end{figure}

Figures \ref{N-plot1} and \ref{N-plot2} show distributions of fitted values
of $1/p$ as a function of the number of drift chambers used for a track
momentum of 50 GeV$/c$. Results are presented for momentum determination
using only a single view in the drift chamber, and for fits that combine
both views. Table \ref{N-table} summarizes chamber number dependence of
fractional resolution and tracking efficiency.

\begin{table}[tbp] \centering%
\begin{tabular}{|l|l|l|l|l|}
\hline
${\bf N}${\bf (chambers)} & 7 & 14 & 28 & 42 \\ \hline
${\bf \sigma /p}^{\mbox{1 view}}$ & 0.56 & 0.49 & 0.30 & 0.20 \\ \hline
${\bf \sigma /p}^{\mbox{1 view}}${\bf (pred)} & 0.52 & 0.37 & 0.27 & 0.22 \\ 
\hline
${\bf \sigma /p}^{\mbox{2 views}}$ & 0.44 & 0.35 & 0.22 & 0.16 \\ \hline
$\sigma /p^{\mbox{2 views}}${\bf (pred)} & 0.37 & 0.26 & 0.19 & 0.15 \\ 
\hline
${\bf \varepsilon }^{\mbox{1 view}}{\bf (\%)}$ & 65 & 91 & 99 & 97 \\ \hline
${\bf \varepsilon }^{\mbox{2 views}}{\bf (\%)}$ & 32 & 82 & 99 & 95 \\ \hline
\end{tabular}
\caption{
Dependence of resolution, ($\sigma_p\over{p}$)
and reconstruction efficiency ($\epsilon$)  on number of chambers used
in the fit for 50 GeV$/c$ input muons.  Resulted are presented for
fits using one or both views of the drift chamber.
The Monte Carlo sample
contained $10^4$ events, so statistical errors are of order 1\%.
\label{N-table}}%
\end{table}%

Agreement between observed resolution from the full reconstruction and Eq. 
\ref{MCS P-error} is satisfactory at 50 and 100 GeV$/c$. At 20 GeV$/c$, the
observed resolution is considerably better than the prediction. Energy loss
in the target is a significant fraction of the total muon energy in this
case. The fitting procedure incorporates energy loss effects. Their
inclusion introduces a second source of correlation between longitudinal
chamber position and momentum that evidently enhances the resolution.
Resolution at 200 GeV$/c$ is about $30\%$ worse in the Monte Carlo than
predicted. The source for this disagreement is not fully understood,
although it may be related to the significant tail that occurs on the high
momentum side for fits to very high energy muons tracks. There are also
small biases evident in the momentum reconstruction that, while much less
than the momentum resolution, are not yet understood. For 50 GeV$/c$ muons,
the resolution is observed to scale with the number of chambers as $1/\sqrt{N%
}$, in agreement with the prediction. The expected $\sqrt{1/2}$ improvement
in resolution when combing the independent $x$ and $y$ views is also
observed.

\section{Systematic Errors}

A proper survey of systematic errors requires treatment of real data, which
will be presented in a forthcoming publication. A few obvious sources are
commented upon here.

The theory of multiple Coulomb scattering is well-established\cite{Moliere,
Bethe, Scott}. The critical parameter that enters into fitting is $\mu _k$,
which represents the effective rms of a Gaussian approximation to the
distribution of the projected scattering angle. As discussed in Appendix \ref
{MCS parameter}, different estimates for $\mu _k$ agree to within $\sim 2\%$
using simple parametrizations. It seems likely that this error could be
reduced to negligible levels if a careful application of the Moliere theory
is applied to a single material.

Any drift chamber misalignment will produce a bias in the MCS\ fitting
procedure that produces a fitted momentum estimate that is systematically
lower than the true value. Misplaced chambers will effectively introduce
extra scatter between hits on a track. The only way the fitting routine can
account for the extra scatter is to lower the momentum, thus increasing the
contribution of MCS to ${\bf V}(p)$. If the misalignment is random, then the
MCS fit will return relatively poor (high) values of the likelihood
function. However, a correlated mis-alignment can mimic the effects of MCS\
fairly well. It is thus critical to have an accurate tracking chamber
alignment.

Spurious hits not directly associated with the muon track from chamber noise
or multiple-pulsing of electronics will also produce undesirable scatter
that will bias momentum fits towards too low values. Electromagnetic shower
particles produced by high energy particles in dense calorimeters will
produce similar effects. It is thus important to use only hits from
``quiet'' intervals of the muon track where there is no possible ambiguity.

As is evident in Fig. \ref{sigma calc} the resolution on the MCS fit is a
fairly strong function of the chamber resolution $\sigma _0$. The MCS\
momentum estimate will also be biased by incorrect $\sigma _0$ values in a
positively correlated way, i.e., the fit will return too high values of
momentum to reduce the MCS\ error contribution if $\sigma _0$ is input at a
value larger than its true value.

\section{Conclusions}

It has been demonstrated that momentum estimates based on multiple Coulomb
scattering can be extended to work for very high energy muons in dense
calorimeters instrumented with a sufficient number of typical drift chamber
tracking detectors. Using the MCS\ technique could allow high energy
neutrino experiments such as NuTeV\ at Fermilab to increase their acceptance
for low energy, wide angle muons that exit their detector before a
spectrometer momentum measurement is possible. Other possible applications
exist in large detectors being assembled for long baseline neutrino
oscillation searches, such as MINOS\cite{minos} at Fermilab.

\appendix 

\section{\label{NuTeV Detector}The NuTeV Detector}

The NuTeV experiment will be used to provide specific examples of MCS
momentum determination using analytic calculations and a detailed Monte
Carlo simulation. NuTeV employs the Fermilab Lab E neutrino detector\cite
{Lab E Calorimeter,Lab E Spectrometer} in a newly constructed high intensity
sign-selected neutrino beam\cite{SSQT}. The experiment is currently taking
data with the primary goal of measuring the weak mixing angle $\sin ^2\theta
_W\equiv 1-M_W^2/M_Z^2$ to a factor of 2-3 times the precision achieved in
previous neutrino experiments. NuTeV will also make improved measurements of
nucleon structure functions and will perform high sensitivity searches for
processes not predicted by the Standard Model of electroweak physics.

The detector consists of a 690 ton iron target-calorimeter followed by an
iron-core toroid spectrometer. The target-calorimeter is a 3 m$\times $3 m$%
\times $18 m volume centered with its long side parallel to the neutrino
beam. Hadron energy measurement, longitudinal event vertex determination,
and event triggering and timing are performing using eighty-four 3 m$\times $%
3 m $\times $2.5 cm liquid scintillation counters read out through
wavelength-shifter bars to phototubes on each of the four corners of the
counter. The scintillation counters are separated by 10 cm of steel, and
provide a sampling-dominated hadronic energy resolution of $\sigma _E/E=0.9/%
\sqrt{E}+0.3/E$.

Transverse event vertex position and muon track angle in charged current
events are measured in the target-calorimeter using up to forty-two 3m$%
\times $3m drift chambers which are spaced every 20 cm of steel. The drift
chambers consist of two orthogonal planes, each containing twenty-four 12.7
cm wide cells. Each cell has two sense wires, permitting local left-right
position ambiguity resolution. The chambers use a mixture of $50\%$ argon- $%
50\%$ ethane that produces a uniform 50 $\mu $m/ns drift velocity. Both
sense wires of each cell are connected through shaping pre-amplifiers to
multi-hit TDC's; the TDC's have 4 ns time buckets and can buffer up to 32
hits. When constructed and initially tested in the early eighties, the drift
chamber resolution was measured to be 225 $\mu $m; the current resolution is
estimated for studies in this paper to be $\sigma _0\simeq 500$ $\mu $m.
Counting support structure, scintillation oil and other passive detector
components, the drift chambers are separated by $42.4$ cm, corresponding to $%
12.2$ radiation lengths.

The toroid spectrometer the follows the calorimeter provides a $10\%$ MCS
momentum resolution for muons from $5-500$ GeV$/c$.

NuTeV detector response is simulated using a Geant-based package\cite{Geant}
that produces output in the same form as the on-line data acquisition
system. Multiple scattering is simulated using the default Moliere
scattering option of Geant. Muon energy loss in iron is simulated using the
Landau fluctuation process for restricted energy loss. Important
contributions from catastrophic energy loss from muon bremsstrahlung and $%
e^{+}e^{-}$ pair emission are carefully modelled by lowering Geant photon
and electron energy tracking cuts to 0.1 and 1.0 MeV, respectively.

\section{\label{Resolution}Derivation of MCS Momentum Resolution}

If one ignores energy loss effects, the track error matrix can be expressed
as 
\begin{eqnarray}
{\bf V}(p) &=&\sigma ^2{\bf I}+\frac{p_0^2}{p^2}{\bf S}(p_0^2) \\
&=&{\bf V}(p_0)+\left( \frac{p_0^2}{p^2}-1\right) {\bf S}(p_0).  \nonumber
\end{eqnarray}
If 
\begin{equation}
{\bf V}(p_0)\left| n\right\rangle =\lambda _n^2\left| n\right\rangle ,
\end{equation}
i.e., $\left\{ \lambda _n^2\right\} $ are the eigenvalues of ${\bf V}(p_0)$,
then 
\begin{equation}
{\bf S}(p_0)\left| n\right\rangle =\left( \lambda _n^2-\sigma ^2\right)
\left| n\right\rangle ,
\end{equation}
and thus, 
\begin{equation}
{\bf V}(p)\left| n\right\rangle =\left[ \frac{p_0^2}{p^2}\left( \lambda
_n^2-\sigma ^2\right) +\sigma ^2\right] \left| n\right\rangle ,
\end{equation}
so that 
\begin{equation}
\log (\det {\bf V}(p))=\sum_n\log \left[ \frac{p_0^2}{p^2}\left( \lambda
_n^2-\sigma ^2\right) +\sigma ^2\right] .
\end{equation}
Letting $p=p_0+\delta _p:$%
\begin{eqnarray}
\log (\det {\bf V}(p)) &\simeq &\sum_n\log \left( \lambda _n^2\right) -\frac{%
2\left( \lambda _n^2-\sigma ^2\right) }{\lambda _n^2}\left( \frac{\delta _p}{%
p_0}\right)  \label{log-det} \\
&&+\frac{\left( \lambda _n^2-\sigma ^2\right) \left( \lambda _n^2+2\sigma
^2\right) }{\lambda _n^4}\left( \frac{\delta _p}{p_0}\right) ^2.  \nonumber
\end{eqnarray}

Expanding the $\chi ^2$ function yields, for a linear fit, only one momentum
dependence that remains after averaging:

\begin{eqnarray}
\left\langle \frac 12\frac{\partial \chi ^2(\theta _0,y_0,p_0)}{\partial
p\partial p}\delta _p\delta _p\right\rangle &=&\frac 12\left\langle \left( 
\vec{y}-\theta _0\vec{z}_1-y_0\vec{z}_0\right) ^T\frac{\partial ^2{\bf V}%
(p)^{-1}}{\partial p\partial p}\right. \\
&&\left. \times \left( \hat{y}-\theta _0\vec{z}_1-y_0\vec{z}_0\right)
\right\rangle \delta _p\delta _p  \nonumber \\
&=&\frac 12Tra\left( \frac{\partial ^2{\bf V}(p)^{-1}}{\partial p\partial p}%
{\bf V}(p)\right) \delta _p\delta _p.  \nonumber
\end{eqnarray}
The trace can be expanded in the eigenvalues of ${\bf V}(p_0):$%
\begin{eqnarray}
Tra\left( \frac{\partial ^2{\bf V}(p)^{-1}}{\partial p^2}{\bf V}(p)\right)
&=&\sum_{n,m}\left\langle n\right| \frac{\partial ^2{\bf V}(p_0)^{-1}}{%
\partial p^2}\left| m\right\rangle \left\langle m\right| {\bf V}(p_0)\left|
n\right\rangle  \label{Trace V} \\
&=&\sum_n\frac{2\left( \lambda _n^2-\sigma ^2\right) \left( \lambda
_n^2-4\sigma ^2\right) }{p_0^2\lambda _n^4}.  \nonumber
\end{eqnarray}

Combining Eq. \ref{log-det} with Eq. \ref{Trace V} yields 
\begin{equation}
\frac{\partial ^2{\cal L}}{\partial \left( \delta _p\right) ^2}=2\sum_n\frac{%
\left( \lambda _n^2-\sigma _0^2\right) ^2}{p_0^2\lambda _n^4}.
\end{equation}
For a linear fit, there are no correlation terms between $p$ and $\theta _0$
or $p$ and $y_0$, so one has directly 
\begin{equation}
\frac{p_0^2}{\sigma _p^2}=2\sum_n\frac{\left( \lambda _n^2-\sigma
_0^2\right) ^2}{\lambda _n^4}.
\end{equation}
The inverse of the squared fractional momentum error is twice the sum of the
ratios of squared eigenvalues of the MCS matrix to the total error matrix.

One can alternatively write 
\[
\lambda _n^2-\sigma _0^2=\frac{\mu ^2}{p^2}\xi _n^2, 
\]
where $\left\{ \xi _n^2\right\} $ are the eigenvalues of the dimensionless
reduced scattering matrix 
\[
{\bf \tilde{S}\ =}\frac{p^2}{\mu ^2\Delta ^2}{\bf S}(p), 
\]
yielding equation \ref{MCS P-error}.

\section{\label{MCS parameter}MCS\ Scattering Parameter}

The critical parameter in the error matrix is the quantity $\mu _k$
appearing in Eq. \ref{MCS angle}. The conventional PDG parametrization\cite
{Highland,Lynch and Dahl} 
\begin{equation}
\mu _k=0.0136\sqrt{\frac{\Delta _k}{X_k}}\left( 1+0.038\ln (\Delta
_k/X_k)\right) ,  \label{MCS angle}
\end{equation}
is the effective $\sigma $ of a Gaussian approximation to that part of the
exact Moliere distribution that encompasses $98\%$ of scattering angles. The
NuTeV drift chambers are separated by $12.2\pm 0.1$ radiation lengths,
yielding a value of $\mu _k$ of $0.0520\pm 0.0002$ GeV$/c$ for single
chamber separation. This value differs by $(-0.7+3.4\ln n)\%$ from the
still-used simpler expression $\mu _k=0.015\sqrt{\Delta _k/X_0}$, where $n$
is the separation between hits in numbers of chambers. The difference is
small for hits separated by a single chamber, but becomes non-negligible for
cases where multi-chamber gaps between hits develop due to inefficiency and
noise.

Lynch and Dahl\cite{Lynch and Dahl} have analyzed the validity of Eq. \ref
{MCS angle} and give an alternate and more accurate parametrization in terms
of the Moliere characteristic angle $\chi _c$ and screening angle $\chi
_\alpha $, given by 
\begin{eqnarray}
\left( \frac{p_k}{1\mbox{ GeV}/c}\right) \chi _c(k) &=&\left( 4.0\times
10^{-4}\right) \sqrt{Z_k\left( Z_k+1\right) \rho _k\Delta _k/A_k}, \\
\left( \frac{p_k}{1\mbox{ GeV}/c}\right) \chi _\alpha (k) &=&\left(
4.48\times 10^{-6}\right) \sqrt{Z_k^{2/3}\left( 1+3.34Z_k^2\alpha ^2\right) }%
,
\end{eqnarray}
where $Z_k$ and $A_k$ are the atomic number and atomic weight of the
material in the scattering region, $\rho _k\Delta _k$ is the thickness of
the region in g-cm$^2$, $p_k$ is the momentum of the charge 1$e$
ultra-relativistic particle, and $\alpha $ is the fine-structure constant.
With these definitions, 
\begin{equation}
\left( \frac{\mu _k}{p/\left( \mbox{1 GeV}/c\right) }\right) ^2=\frac{\chi
_c^2(k)}{1+F^2}\left( \frac{1+v(k)}{v(k)}\ln \left( 1+v(k)\right) -1\right) ,
\end{equation}
where $v(k)=0.5\Omega (k)/(1-F)$, $\Omega (k)=\chi _c^2(k)/\chi _\alpha
^2(k) $ is the mean number of scatters, and $F$ is the fraction of scatters
defined in a Gaussian fit to the Moliere distribution. Using Lynch and
Dahl's prescription for including small effects of non-steel components of
the NuTeV target yields, for $F=0.98$, $\mu _k=0.0528$ GeV$/c$-for one
chamber separation in NuTeV, $1.9\%$ higher than the PDG formula prediction.

\section{\label{Fit Procedure}MCS Fitting Procedure}

\subsection{Hit Selection}

Initial track selection is performed using NuTeV's conventional calorimeter
track finding and reconstruction software which has been tested over a
period of many years in dozens of physics analyses. Inclusion of hits not
associated with the muon track bias the MCS momentum fit by pulling the
momentum to low values to handle the anomalously large scatter. These hits
can originate from the hadron shower induced by the neutrino interaction,
from delta rays and other electromagnetic shower components produced by
muon, and from electronic noise. To minimize their effect, only drift
chamber cells with a single hit are used and a local trajectory requirement
is imposed. The latter condition requires a hit position in a given drift
chamber to be within 1 mm of the average position of hits from immediately
neighboring chambers. One mm is about three times the typical single chamber
MCS displacement for a 10 GeV$/c$ muon.

\subsection{Energy Loss Correction}

A 50 GeV$/c$ muon passing completely through the target calorimeter loses $%
\sim 17$ GeV of energy on average. This degradation of energy causes
downstream hits to spread further than would be predicted by the MCS error
matrix calculated with a single momentum. This effect is taken into account
by assuming an average energy loss of $\left\langle dp/dz\right\rangle
=0.0249$ GeV/cm of detector, as calculated from the Geant simulation. No
attempt is made to correct on an event-by-event basis for catastrophic
energy loss, although such a procedure could be developed using the target
scintillation counters.

A smaller issue is the momentum $p_k$ that should be used in Eq. \ref{MCS
matrix}. A simple analysis shows that this value is the geometric mean of
the momentum evaluated at the beginning and end of scattering medium $k$ 
\begin{equation}
p_k=\sqrt{\left[ p_0-\left\langle \frac{dp}{dz}\right\rangle \left(
z_{k-1}-z_0\right) \right] \left[ p_0-\left\langle \frac{dp}{dz}%
\right\rangle \left( z_k-z_0\right) \right] }.
\end{equation}

\subsection{Fitting Algorithm}

Events with at least five drift chamber hits in at least one view are input
to a likelihood minimization routine. The routine operates iteratively,
varying the momentum from iteration to iteration, but holding $p_0$ fixed in
the minimization of $\chi ^2(\theta _0,y_0,p_0)$ with respect to $\theta _0$
and $y_0$. The momentum is set to 25 and 50 GeV/$c$ for the first two
iterations, and then minimized by approximating ${\cal L}$ as a quadratic
function of $\log p$. A fit typically converges in 8-10 iterations. A
straight-through muon with 42 hits in each of two views requires about one
second of CPU time on a 200 MHz Intel workstation.

\begin{figure}[tbph]
\psfig{file=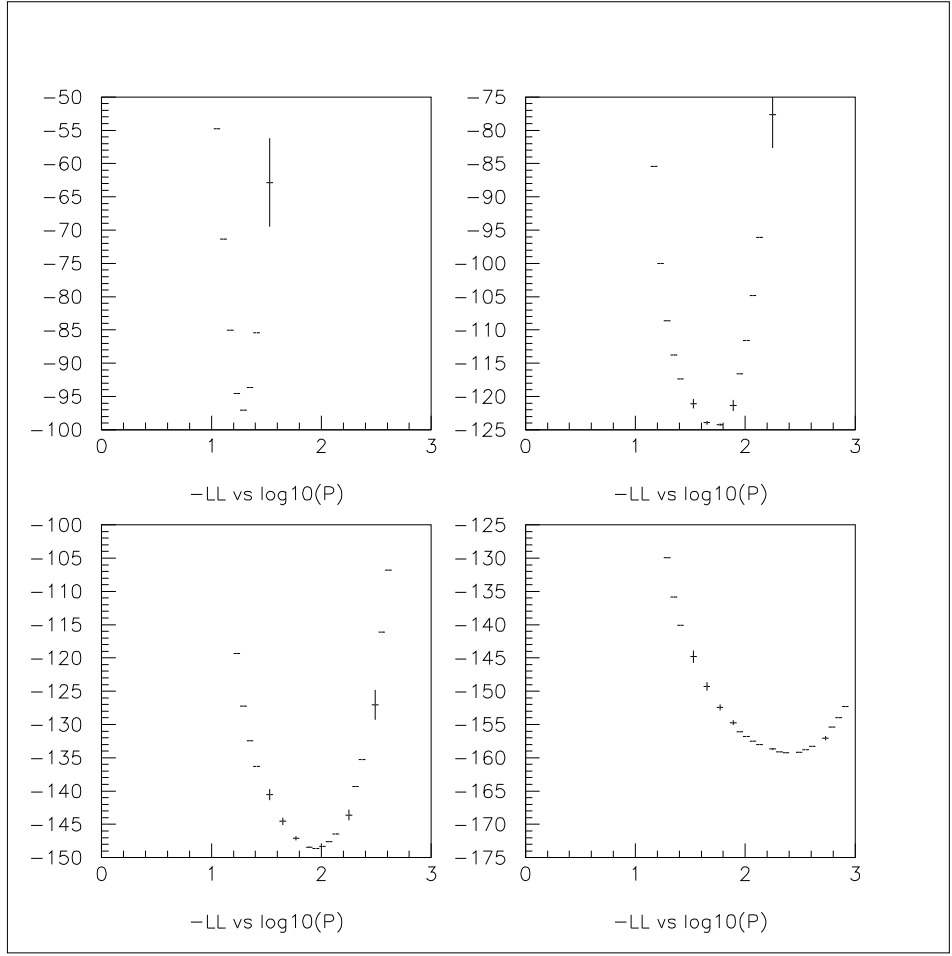,clip=,height=5.0in}
\caption{ Typical likelihood function profiles plotted 
vs $\log{p}$ for (clockwise
from upper left) 20 GeV$/c$, 50 GeV$/c$, 200 GeV$/c$, and 100 GeV$/c$ muons
fitted using 42 drift chambers in the NuTeV geometry. }
\label{log-like-P}
\end{figure}

\begin{figure}[tbph]
\psfig{file=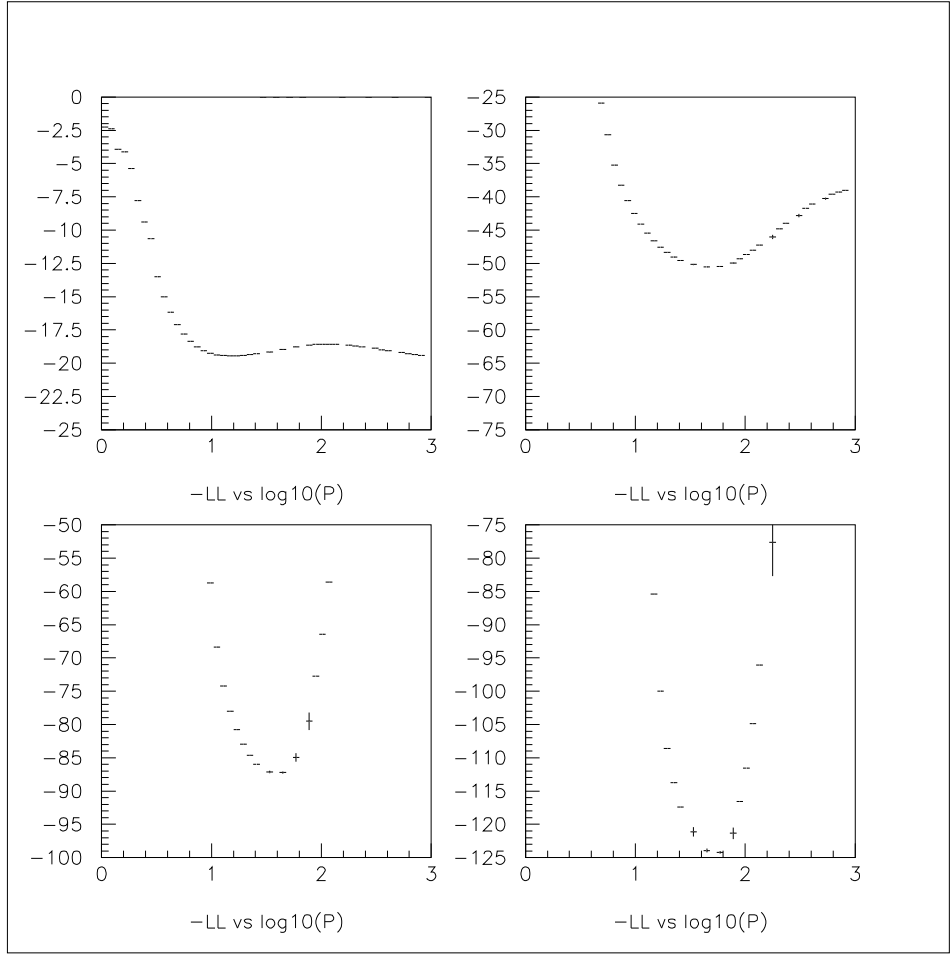,clip=,height=5.0in}
\caption{ Typical likelihood function profiles plotted 
vs $\log{p}$ for 50 GeV$/c$
muons fitted using (clockwise from upper left) 7, 14, 42, and 28 drift
chambers in the NuTeV geometry. }
\label{log-like-N}
\end{figure}

Figure \ref{log-like-P} show the likelihood function profile for typical
fits to 20 GeV$/c$, 50 GeV$/c$, 100 GeV$/c$, and 200 GeV$/c$\ momentum muons
using 42 drift chambers. The shape of the function is approximately
parabolic for a large region about the minimum when plotted vs $\log p$.
Figure \ref{log-like-N} shows ${\cal L}$ vs $\log p$ profiles for 50 GeV$/c$
muons with 7, 14, 28, and 42 drift chambers used in the fit. As the number
of chambers becomes small, the ${\cal L}$ distribution becomes skewed with a
marked tail at high momentum. This behavior is expected since the relatively
small and poorly determined scatter that occurs among a small number of
chambers can be consistently described by nearly any momentum value higher
than the true value. Note that ${\cal L}$ rises steeply for momentum values
lower than the true value even for a small number of chambers. The MCS
method may thus be used to set a useful lower bound on track momentum even
if a reliable estimate of the true value is not feasible.

\end{document}